\documentclass{article}
\usepackage{amsfonts, amssymb, amsmath, amsthm, mathrsfs, chet, cite}
\usepackage{graphicx} % Required for inserting images
\usepackage{braket}
\usepackage{bm}% bold math\
\usepackage{cancel}

\usepackage{subcaption}  % To create subfigures
\usepackage{algorithm} 
\usepackage{algpseudocode}
\usepackage{xcolor}
\usepackage{array}
\usepackage{tabularx}
\usepackage{braket}

\begin{document}

\title{How many qubits does a machine learning problem require?}
\author{Sydney Leither, Michael Kubal, Sonika Johri \email{sjohri@coherentcomputing.com}}
\affiliation{Coherent Computing Inc.}

\abstract{
For a machine learning paradigm to be generally applicable, it should have the property of universal approximation, that is, it should be able to approximate any target function to any desired degree of accuracy. In variational quantum machine learning, the class of functions that can be learned depend on both the data encoding scheme as well as the architecture of the optimizable part of the model. Here, we show that the property of universal approximation is constructively and efficiently realized by the recently proposed bit-bit data encoding scheme. Further, we show that this construction allows us to calculate the number of qubits required to solve a learning problem on a dataset to a target accuracy, giving rise to the first resource estimation framework for variational quantum machine learning. We apply bit-bit encoding to a number of medium-sized classical benchmark datasets and find that they require only $27$ qubits on average for encoding. We extend the basic bit–bit encoding scheme to a variant that efficiently supports batched processing of large datasets. As a demonstration, we apply this new scheme to subsets of a giga-scale transcriptomic dataset. This work establishes bit-bit encoding not only as a universally expressive quantum data representation, but also as a practical foundation for resource estimation and benchmarking in quantum machine learning.
}

\maketitle

\section{Introduction}
The use of quantum computers for machine learning and artificial intelligence applications is a relatively new research area that could provide a means to model datasets which contain correlations that are hard to model classically. Quantum machine learning algorithms have been tested on near-term quantum hardware for various real-world datasets \cite{Gujju24}, \cite{elton_copula}, \cite{zhu2022copulabased}, \cite{iaconis2023tensor}, \cite{Johri20}, \cite{Rudolph20}, \cite{cherrat2022quantum}, \cite{silver2023mosaiq}, \cite{Silver_Patel_Tiwari_2022}. In particular, variational quantum machine learning, which uses parameterized unitaries that can be trained to model the correlations in a dataset analogous to classical neural networks \cite{Wang_2024}, can be applied to a wide variety of problems. However, compared to other applications such as chemistry~\cite{Feynman1982}, \cite{Lloyd1996},\cite{AspuruGuzik2005},\cite{Reiher2017} or cryptography~\cite{Shor1997}, where algorithms with well-understood quantum advantage and extensively studied resource estimation are known, there are many more open questions around the use of quantum computers to learn from data.

Among the fundamental questions is one of expressivity --- how does one design a variational quantum ansatz, which we refer to as a quantum model, that can approximate the underlying function defined by a given dataset? To answer this question meaningfully, we must note that real-world datasets seldom follow theoretically tractable rules. Therefore, the quantum model should have the property of \emph{universal approximation} --- that is, it should be able to approximate any function with any desired degree of accuracy. Further, the construction which realizes the universal approximation should not itself require exponentially scaling computation, either quantum or classical. If a quantum learning model does not have the property of universal approximation, it is likely not useful in real life. Even if such a model works in a limited setting, we cannot expect it to give good results in the limit of large and complex datasets. Indeed, in classical machine learning, universal approximation theorems \cite{Cybenko1989, Hornik1991} provide the mathematical justification for using neural networks for large and complex datasets.

Schuld et al.~\cite{schuld2021effect} showed that the expressivity of a quantum model is constrained by the method used for encoding data into a quantum state. Specifically, a quantum model can be expanded as a partial Fourier series in the encoded data where the frequencies present in the series are set by linear combinations of the eigenvalues of the encoding operators. Following this, it is possible to increase the number of frequencies in the series by reusing the data multiple times in the quantum ansatz, a technique known as data re-uploading \cite{perez2020data}. The time for loading is then lower-bounded by the quantity of information to be loaded multiplied by the number of times it is loaded. With this method, even an image from a simple dataset like MNIST, which has 784 pixels, is practically expensive to load directly into a quantum state. Since this limit is fundamental, it is clear that some kind of compression is required before loading, for which techniques like principal component analysis are often used. However, it remains unclear how much the degree of compression and the number of data re-uploadings affect the accuracy that can be reached by the model.
% However, as we will show, this still does not create universal approximation since there is no constructive method for approximating the coefficients of the Fourier terms to an arbitrary degree of precision.

In this work, we show that the bit-bit encoding technique \cite{johri2025bitbitencoding}, which was recently proposed for supervised quantum learning, directly addresses the problem of achieving efficient compression with predictable accuracy. In this encoding, both the input and the output are encoded as binary strings following a dimensionality reduction procedure. These binary strings correspond to the basis states of a quantum computer at initialization and measurement. For a small number of qubits, there are collisions in which the same input corresponds to more than one output, but as the number of qubits increases, the encoding becomes more precise. In this limit, we prove that a variational quantum model with bit-bit encoding has the property of universal approximation. Further, the classical pre-processing required before data loading utilizes techniques such as principal component analysis which scale only polynomially in the dataset size.

A useful emergent property of the bit-bit encoding construction is that it is possible to efficiently calculate the number of qubits required to encode a dataset to a desired degree of accuracy. This then gives us the first framework for \emph{resource estimation} in quantum machine learning. Using this framework, we can, for the first time, estimate the number of qubits required to successfully train a model on a given dataset using a variational approach. In turn, this analysis lets us assess the potential for quantum advantage on a dataset as we can calculate whether a learning problem fits into a number of qubits that is classically simulatable.

We introduce $Q_\text{dataset}(x)$, a resource metric that estimates the number of qubits required to model a dataset to accuracy $x$ using bit-bit encoding; $Q_\text{dataset}(1.0)$ is the number of qubits required to achieve both $100\%$ training and testing accuracy. Starting with the MNIST dataset, we illustrate how $Q_\text{dataset}(x)$ is derived by incrementally increasing qubit allocations until collisions on both the train and test datasets vanish, revealing a trade-off between maximizing theoretical training accuracy and achieving generalization. We find that medium-sized classical benchmark classification datasets have a $Q_\text{dataset}(1.0)$ of $27$ on average. Comparisons across dimensionality reduction schemes show that bit-bit encoding achieves more efficient data compression with schemes that produce independent or de-correlated features. Finally, scaling experiments on a transcriptomic dataset reveal examples where subsets of the dataset have a much higher $Q_\text{dataset}(1.0)$ than the benchmark datasets despite having comparable dimensionality, indicating that large transcriptomic dataset learning problems are promising candidates for quantum advantage.

We note that we only consider classical datasets in this work. Although there has been progress in quantum machine learning for quantum data \cite{Nagano2023QuantumDataLearning}, \cite{Huang2022QuantumAdvantageExperiments}, practical use cases remains limited. In contrast, quantum machine learning applied to classical data offers a far broader range of potential applications.

The paper is organized as follows. We begin with an overview of bit-bit encoding in Section 2, followed by an analysis of expressive power and universal approximation in commonly used encoding schemes in Section 3. Section 4 presents calculations of $Q_\text{dataset}$ on classical benchmark datasets across dimensionality reduction schemes along with an extension of bit-bit encoding for large-scale data encoding. Section 5 demonstrates training of a classifier on a bit-bit encoded dataset. We conclude with a discussion in Section 6.

%Finally, we point readers to an online tool from Coherent Computing that you can use to estimate the number of qubits required for classification on your dataset.

\section{Methods}
\subsection{Overview of Bit-Bit Encoding} \label{sec:bbe}
Bit-bit encoding \cite{johri2025bitbitencoding} encodes datasets with numeric features and discrete class labels into bitstrings, which allows the dataset to be loaded into the Hilbert space of a given number of qubits for classification. Consider a supervised learning problem for a dataset with a feature matrix $X \in \mathbb{R}^{s \times n}$ and class label vector $y \in \{1, 2, \ldots, c\}^s$, where $s$ is the number of samples, $n$ is the number of features, and $c$ is the number of class labels. For a classification problem, the learning task consists of finding an efficiently computable function that maps $X$ to $y$. Typically, the function is many-to-one, that is, $c<|X|$.

Bit-bit encoding converts the feature matrix $X$ into a discretized feature matrix $Z$, compactly encoded into $N_x$ qubits. Bit-bit encoding works as follows. First, the dataset is split into train and test sets: $X_{\text{train}}$, $y_{\text{train}}$, $X_{\text{test}}$, and $y_{\text{test}}$. A dimensionality reduction model, principal component analysis (PCA) by default, is fit on $X_{\text{train}}$ and applied to $X_{\text{test}}$, reducing both sets to $D$ components. The number of bits assigned to each component is calculated with $$b_d=\lfloor N_x\cdot \frac{I(X_d;y)}{\sum_{i=1}^D{I(X_i;y)}} \rceil$$ where $I$ is the mutual information and $N_x$ is the number of bits allocated to the discretized feature matrix. The dimension-reduced $X_{\text{train}}$ and $X_{\text{test}}$ are then min-max normalized so that each sample lies in $[0,1]^D$. A copula transformation is fit on the reduced and normalized $X_{\text{train}}$ and applied to the reduced and normalized $X_{\text{test}}$, mapping each component to its empirical cumulative distribution function so that the feature distributions are approximately uniform. Finally, to encode the reduced, normalized, and transformed $X_{\text{train}}$ and $X_{\text{test}}$, each data point $X_{sd}$ (the $s^{th}$ sample of the $d^{th}$ feature) is discretized into $b_d$ bits of precision with $$Z_{sd}=\lfloor X_{sd}2^{b_d}\rfloor$$ where $Z$ is the discretized feature matrix. As the number of qubits $N_x$ grows, $Z$ provides a more precise representation of the original dataset.

Discretizing the feature matrix also compresses it, potentially creating collisions at the input as a particular feature vector $z \in Z$ may occur more than once, each time corresponding to the same or different value of $y_s \in y$. The samples in the discretized feature matrix can then be considered as samples from a joint probability distribution $f(Z=z, y=y_s)$ over correlated random variables $(Z, y)$. For the purpose of classification, the ``correct'' output is then taken to be the mapping which occurs most frequently, that is, $$C(z) = \arg \max_{y_s} f(z, y_s)$$ is the classification function to be learned. $Z$ can be exactly loaded into a quantum state that lives in the Hilbert space of $N_x$ qubits using Pauli-X gates acting on the $\ket{0}$ state to create the computational basis state $\ket{z}$. Similarly, the class labels can be mapped onto the computational basis state $\ket{y}$ of $N_y = \lceil{\log_2{c}}\rceil$ qubits. Thus the classification function is of the form $C: \{0, 1\}^{N_x} \to \{0, 1\}^{N_y}$.

Given this form of the classification function, the quantum learning problem can be cast in terms of an operator $U_*$ that has the following action on $N_q=N_x+N_y$ qubits:
\begin{equation} \label{eq:bit_bit_unitary}
    \ket{0}\ket{z}\xrightarrow{U_*} \ket{C(z)}\ket{g(z)}
\end{equation}
where the states $\ket{g(z)}$ and $\ket{z}$ are defined over $N_x$ qubits and the state $\ket{0}$ defined over $Q_y$ qubits is mapped by $U_*$ to the state $\ket{C(z)}$. The state $\ket{g(z)}$ can be disregarded. As shown in Section~\ref{sec:universality}, $U_*$ is unitary when $C(z)=C(z')=>\langle g(z) | g(z')\rangle =\delta_{z,z'}$.

The learning task is that of finding a unitary $U$, parameterized by variables $\theta$, that maximizes the probability of measuring $\ket{C(z)}$ when acting on $\ket{0}\ket{z}$. This task is defined as
\begin{align}
    U(\vec{\theta})\ket{0}\ket{z}=\sum_k \sqrt{P_{k,z}(\vec{\theta})}e^{i\phi_{k,z}(\vec{\theta})}\ket{k}\ket{g_{k,z}(\vec{\theta})}.
\end{align}
A linear loss function can be constructed as
\begin{align} \label{eq:loss}
    \bar{L}(\vec{\theta})=1-\sum_{z\in \mathcal{D}_{\vec{b}}} f(z) P_{C(z),z}(\vec{\theta}),
\end{align}
where $0\leq \bar{L}\leq 1$ and $f(z)$ is the frequency of occurrence of $z$ in the dataset.  This loss function returns the probability of the model giving the incorrect answer when queried. Similarly, non-linear loss functions can also be constructed. If $U(\vec{\theta})=U_*$ and there is no noise, even one shot would suffice to classify the input data sample.

We distinguish the bit-bit encoding technique from a simple basis encoding as it can be used to represent datasets that do not originally have a binary representation, but are converted into binary representation through a process of dimensionality reduction followed by mutual-information based discretization. Indeed, a basis encoding that utilized only a naive discretization of the dataset would require many more qubits. For instance, basis encoding a dataset consisting of molecular fingerprint samples \cite{boy2025encodingmolecularstructuresquantum}, which are naturally binary, would take thousands of qubits.

\section{Universal Approximation}
Next, we examine the expressive ability of bit-bit encoded quantum models and make a comparison with other encoding schemes. In this section, we discuss universal approximation for quantum machine learning and show that it can be straightforwardly achieved with bit-bit encoding whereas it is harder to construct efficiently for amplitude and angle encoding, the two other commonly used encoding schemes. 

\emph{Definition:} Let $\mathcal{X} \subseteq \mathbb{R}^n$ and $\mathcal{Y} \subseteq \mathbb{R}^k$ be compact sets.  
Consider a family of parameterized quantum models 
\[
\mathcal{F} = \left\{ f_{\theta} : \mathcal{X} \to \mathcal{Y} \,\middle|\, 
f_{\theta}(x) = 
\operatorname{Tr}\!\left( M \, U(\theta, x)\, 
\rho_{\mathrm{in}}\, U(\theta,x^\dagger \right),
\;
\theta \in \Theta \subseteq \mathbb{R}^p \right\},
\]
generated by preparing an input state $\rho_{\mathrm{in}}$, applying a parameterized quantum circuit $U(\theta,x)$, and measuring with a POVM $M$.

We say that $\mathcal{F}$ is \emph{universally approximating} if, for every continuous function $g : \mathcal{X} \to \mathcal{Y}$ and every $\varepsilon > 0$, there exists $\theta \in \Theta$ such that
\[
\sup_{x \in \mathcal{X}} \, \left\| f_{\theta}(x) - g(x) \right\| < \varepsilon.
\]

\subsection{Universality of bit-bit encoding} \label{sec:universality}
In this section, we show how the bit-bit encoding leads to a universal approximation theorem for classification. First, we prove that the classification operator $U_*$ defined in Eq. \ref{eq:bit_bit_unitary} is unitary.

\textit{Theorem 1: Let $N_x,N_y\in\mathbb{N}$, and set
\[
\mathcal{H}_x := (\mathbb{C}^2)^{\otimes N_x},\qquad
\mathcal{H}_y := (\mathbb{C}^2)^{\otimes N_y},\qquad
\mathcal{H} := \mathcal{H}_y\otimes \mathcal{H}_x .
\]
Let $C:\{0,1\}^{N_x}\to\{0,1\}^{N_y}$ be a classical classifier and
let $\{\ket{g(z)}\in\mathcal{H}_x: z\in\{0,1\}^{N_x}\}$ be unit vectors such that
\begin{equation}\label{eq:reversibility}
C(z)=C(z')\;\Longrightarrow\;\braket{g(z)|g(z')}=\delta_{z,z'} .
\end{equation}
Define a linear map $V$ on the subspace
\[
\mathcal{H}_{\mathrm{in}}:=\mathrm{span}\bigl\{\ket{0^{N_y}}\ket{z}:z\in\{0,1\}^{N_x}\bigr\}\subset\mathcal{H}
\]
by
\begin{equation}\label{eq:spec}
V\bigl(\ket{0^{N_y}}\ket{z}\bigr)=\ket{C(z)}\ket{g(z)}\qquad\text{for all }z\in\{0,1\}^{N_x}.
\end{equation}
Then there exists a unitary $U_*:\mathcal{H}\to\mathcal{H}$ such that
\begin{equation}\label{eq:desired}
U_*\bigl(\ket{0^{N_y}}\ket{z}\bigr)=\ket{C(z)}\ket{g(z)}\qquad\text{for all }z\in\{0,1\}^{N_x}.
\end{equation}}

\textit{Proof:} First, for any $z,z'\in\{0,1\}^{N_x}$,
\begin{align*}
\bigl\langle C(z),g(z)\bigm|C(z'),g(z')\bigr\rangle
&=\braket{C(z)|C(z')}\,\braket{g(z)|g(z')} \\
&=\delta_{C(z),C(z')}\,\braket{g(z)|g(z')} .
\end{align*}
If $C(z)\neq C(z')$ the first factor vanishes; if $C(z)=C(z')$, the condition
\eqref{eq:reversibility} gives $\braket{g(z)|g(z')}=\delta_{z,z'}$. Hence, in all cases,
\begin{equation}\label{eq:ON}
\bigl\langle C(z),g(z)\bigm|C(z'),g(z')\bigr\rangle=\delta_{z,z'}.
\end{equation}
Therefore the set
\[
\mathcal{B}_{\mathrm{out}}:=\bigl\{\ket{C(z)}\ket{g(z)}:z\in\{0,1\}^{N_x}\bigr\}
\]
is an orthonormal family in $\mathcal{H}$. Because $|\{0,1\}^{N_x}|=2^{N_x}$ and
$\dim\mathcal{H}_{\mathrm{in}}=2^{N_x}$, we have
$\dim\mathrm{span}(\mathcal{B}_{\mathrm{out}})=2^{N_x}$. Thus $V$ in \eqref{eq:spec} is an
isometry from $\mathcal{H}_{\mathrm{in}}$ onto
\[
\mathcal{H}_{\mathrm{out}}:=\mathrm{span}(\mathcal{B}_{\mathrm{out}}),
\]
and, because these subspaces have equal dimension, $V$ is unitary between them.
 
We see that condition \eqref{eq:reversibility} enforces reversibility: whenever the classifier $C$ identifies two inputs,
the $g$-register separates them so that the pairs $\bigl(\ket{C(z)},\ket{g(z)}\bigr)$ remain mutually orthonormal.
A standard special case is the reversible oracle $\ket{y}\ket{z}\mapsto\ket{y\oplus C(z)}\ket{z}$,
which corresponds to choosing $\ket{g(z)}=\ket{z}$.

Next, we show that the classification operator $U_*$ can be approximated to any desired degree of accuracy by a quantum circuit constructed out of a finite universal gate set.

\textit{Theorem 2:
Fix integers $N_x,N_y\ge 1$ and let $N_q=N_x+N_y$. For any (deterministic) classifier
$C:\{0,1\}^{N_x}\to\{0,1\}^{N_y}$ there exists a unitary $U_* \in \mathsf{U}(2^{N_q})$ and an
orthonormal family $\{\ket{g(z)}\in (\mathbb{C}^2)^{\otimes N_x}\}_{z\in\{0,1\}^{N_x}}$ such that
\begin{equation}\label{eq:Ustar-action}
U_*\bigl(\ket{0^{N_y}}\ket{z}\bigr)\;=\;\ket{C(z)}\ket{g(z)}\qquad(\forall z\in\{0,1\}^{N_x}).
\end{equation}
Let $\mathcal{G}$ be any finite, fixed, two-qubit universal gate set. Then, for every $\varepsilon>0$, there exists a $\mathcal{G}$-circuit
$U_\varepsilon$ on $N_q$ qubits such that
\begin{equation}\label{eq:SK-approx}
\|U_\varepsilon-U_*\|_{\mathrm{op}}\;\le\;\varepsilon,
\end{equation}
and, consequently, for every $z\in\{0,1\}^{N_x}$,
\begin{equation}\label{eq:state-approx}
\bigl\|U_\varepsilon\ket{0^{N_y}}\ket{z}-\ket{C(z)}\ket{g(z)}\bigr\|\;\le\;\varepsilon.
\end{equation}
In particular, if the first $N_y$ qubits are measured in the computational basis, the outcome
distribution under $U_\varepsilon$ differs from the ideal one
by total variation distance at most $O(\varepsilon)$; hence the implemented classifier coincides
with $C$ on all inputs $z$ except with probability $O(\varepsilon)$.}

\textit{Proof:}
The existence of $U_*$ with the action \eqref{eq:Ustar-action} follows from Theorem 1.

For the approximation claim, by universality of $\mathcal{G}$ the subgroup generated by $\mathcal{G}$
is dense in $\mathsf{SU}(2^{N_q})$. The Solovay--Kitaev theorem guarantees that for any target unitary
$V\in\mathsf{SU}(2^{N_q})$ and any $\varepsilon>0$ there exists a $\mathcal{G}$-circuit
$W_\varepsilon$ whose length is polylogarithmic in $1/\varepsilon$ and such that
$\|W_\varepsilon-V\|_{\mathrm{op}}\le \varepsilon$. Applying this with $V:=U_*$ yields a circuit
$U_\varepsilon$ satisfying \eqref{eq:SK-approx}. Then \eqref{eq:state-approx} is immediate:
$\| (U_\varepsilon-U_*)\ket{0^{N_y},z}\|\le \|U_\varepsilon-U_*\|_{\mathrm{op}}\le \varepsilon$.

Finally, for any projective measurement $M$ (in particular, the computational-basis measurement on
the first $N_y$ qubits), the total variation distance between the outcome distributions arising from
pure states $\ket{\psi}$ and $\ket{\phi}$ is bounded by a constant multiple of
$\|\ket{\psi}-\ket{\phi}\|$; applying this with
$\ket{\psi}=U_\varepsilon\ket{0^{N_y}}\ket{z}$ and $\ket{\phi}=U_*\ket{0^{N_y}}\ket{z}$ gives the stated
$O(\varepsilon)$ misclassification bound.

The two theorems together show that the bit-bit encoding combined with an ansatz that consists of layers of gates from a universal gate set can approximate a classifier to any desired degree of accuracy on a dataset discretized as shown in the previous section. 

Finally, we observe that there exists a maximum number of bits required to encode the samples in a dataset.

\textit{Observation}: \textit{Let $\mathcal{D} = \{(x_i, y_i)\}_{i=1}^s$ denote a dataset, where each feature vector $x_i$ admits a representation using at most $p_x$ bits, and each label $y_i$ admits a representation using at most $p_y$ bits. Then each pair $(x_i, y_i)$ can be uniquely encoded using no more than $p_x + p_y$ bits.}

In fact, the limit where the above observation holds corresponds to a simple basis encoding of the dataset. Further, it is easy to see that as the number of bits increases, the accuracy of the encoding monotonically increases. Thus, as the number of qubits and depth of the ansatz increase, the quantum model can approximate the underlying function expressed by the dataset to any desired degree of accuracy, achieving universal approximation.

% \textit{Corollary: The form of this data encoding presents a universal approximation for any dataset with a finite number of qubits.}

\subsection{Other encoding schemes}
Two other encoding schemes are often utilized in the quantum machine learning literature --- amplitude encoding, and angle encoding. For both schemes, some kind of compression is required in order to make the data practically loadable into the quantum computer. Further, within each scheme, the data can be loaded into the quantum computer multiple times in order to enhance the model's expressivity. However, there does not exist a framework to understand the combined effects of the compression, data re-uploading and design of the variational ansatz on the approximation power of the model.

\textit{Amplitude encoding:} A real input vector $x\in\mathbb{R}^n$ with $\|x\|_2=1$ under amplitude encoding is mapped to the quantum state $|x\rangle=\sum_{i=0}^{n-1}x_i\,|i\rangle\in\mathbb{C}^n$. Applying a parameterized unitary $U(\theta)$ yields the state $|\psi(x,\theta)\rangle=U(\theta)|x\rangle$. Measuring an observable $M$ produces the model output $f(x;\theta)=\langle\psi(x,\theta)|M|\psi(x,\theta)\rangle=\langle x|U^\dagger(\theta)MU(\theta)|x\rangle$. Defining $A(\theta):=U^\dagger(\theta)MU(\theta)$, we obtain $f(x;\theta)=x^\top A(\theta)x$ with $A(\theta)$ Hermitian. Hence the class of functions implementable by amplitude encoding followed by a parameterized unitary and measurement is
\[
\mathcal{F}_{\mathrm{AE}}=\bigl\{f(x)=x^\top A\,x:A\in\mathrm{Herm}(n),\ \|x\|_2=1\bigr\},
\]
namely the set of real quadratic forms on the unit sphere in $\mathbb{R}^n$.
However, $C(S^{n-1})$, the space of continuous functions on the input sphere, is infinite-dimensional, whereas $\mathcal{F}_{\mathrm{AE}}$ is finite-dimensional. Hence $\mathcal{F}_{\mathrm{AE}}$ cannot be dense in $C(S^{n-1})$, meaning amplitude encoding does \emph{not} satisfy universal approximation.

The expressivity of amplitude encoding can be enhanced by loading multiple states in parallel. However, a significant barrier remains in the form of the loading time being proporational to the dimension of the data. Previous work has shown the effectiveness of using tensor network techniques to compress the data before loading it as an amplitude encoded state \cite{iaconis2023quantum, iaconis2023tensor, pollmann_tensor, jumade2023dataloadableshortdepth}, but has not studied the effect of the compression on the model accuracy.

\textit{Angle encoding:} Under angle embedding, each input $x\in\mathbb{R}^n$ is mapped to a product of single-qubit rotations $U_{\mathrm{enc}}(x)=\bigotimes_{j=1}^dR_{\alpha}(\phi_j(x))$ (where $R_{\alpha}$ are Pauli rotations), after which a parametrized ansatz $W(\theta)$ is applied and an observable $M$ is measured, yielding $f(x;\theta)=\langle0|\,U_{\mathrm{enc}}^\dagger(x)W^\dagger(\theta)M W(\theta)U_{\mathrm{enc}}(x)\,|0\rangle$. Writing $A(\theta)=W^\dagger(\theta)MW(\theta)$ and expanding conjugation by $U_{\mathrm{enc}}(x)$ shows that $f$ is a finite trigonometric Fourier series in $x$, namely $f(x;\theta)=\sum_{k}a_k(\theta)\cos(\omega_k\cdot x)+b_k(\theta)\sin(\omega_k\cdot x)$ for some frequencies $\omega_k\in\mathbb{Z}^n$ arising from the Pauli rotation exponents. With $L$ layers of data re-uploading, $|\psi(x,\theta)\rangle=\bigl(\prod_{\ell=1}^LW_\ell(\theta_\ell)U_{\mathrm{enc}}(x)\bigr)|0\rangle$, the attainable frequency set expands to $\Omega_L\subset\mathbb{Z}^n$, so that the expressible function class is the family of trigonometric polynomials $\mathcal{F}_{\mathrm{angle}}^{(L)}=\bigl\{\,f(x)=\sum_{\omega\in\Omega_L}[a_\omega\cos(\omega\cdot x)+b_\omega\sin(\omega\cdot x)]:a_\omega,b_\omega\in\mathbb{R}\,\bigr\}$. 

The set of achievable frequencies $\Omega_L$ generated by nested Pauli rotations depends intricately on the circuit architecture and grows in a highly non-regular, combinatorial fashion.  Consequently, there is no efficient algorithm that, given a dataset $\mathcal{D}$ or a target function $f^\star$, determines the minimum number of re-uploading layers $L$ for which $\mathcal{F}_{\mathrm{angle}}^{(L)}$ contains an $\varepsilon$-approximant to $f^\star$.  Moreover, unlike classical Fourier approximation, where one systematically improves approximation quality by increasing the maximum frequency or degree in a controlled way, the sets $\Omega_L$ are not nested in a structured fashion (for example, $\Omega_{L+1}$ does not simply add all frequencies above a certain cutoff), so increasing $L$ does not provide a principled or monotonic method of reducing approximation error.  Instead, $L$ must be heuristically chosen, trained, and empirically validated, meaning the universal approximation property does not translate into a systematic, constructive approximation procedure.

In line with this, there are several recent papers that aim to enhance the expressivity of angle encoding. \cite{wen2024enhancingexpressivityquantumneural} adds auxiliary qubits to enhance the frequency spectrum of an architecture with angle embedding. \cite{wang2025predictiveperformancedeepquantum} finds that using data re-uploading with circuits that are wider rather than deeper is a more effective strategy. Despite this work on enhancing expressivity, there remains no known method for efficiently and constructively realizing universal approximation with these encoding schemes. 

% In both amplitude and angle encoding, the time taken for a single upload of data is proportional to the dimensionality of the data. As we show in later sections in this paper, the bit-bit encoding does not require a number of qubits that scales with the dimension.

% Indeed, even the number of qubits and observables required for such a model to be universal will scale as powers of $d>1$ \cite{Goto_2021}. 

Combining these results, we conclude that bit-bit encoding provides a direct and efficient route to universality in quantum machine learning models. By construction, every dataset can be represented using a bounded number of qubits, and the corresponding classifier can be realized as a unitary operator acting on the joint input–output space. The Solovay–Kitaev theorem then ensures that such unitaries can be approximated to arbitrary accuracy using a finite universal gate set, thereby establishing universality for bit-bit encoded models. In contrast, there is no practical known construction for amplitude and angle encoding to achieve comparable universality. Thus, bit-bit encoding yields both a conceptually transparent and practically scalable framework for constructing universally approximating quantum models.

% Inadequate performance of repeated uploading

% $\mathcal{F}_{\mathrm{angle}}^{(L)}$ is a finite-dimensional space of trigonometric polynomials and therefore not dense in $C(\mathbb{T}^d)$; consequently, angle embedding does \emph{not} satisfy universal approximation for any fixed $L$, but only in the limit $L\to\infty$, provided the circuit architecture allows $\Omega_L\to\mathbb{Z}^d$, the union $\bigcup_{L=1}^{\infty}\mathcal{F}_{\mathrm{angle}}^{(L)}$ can be dense in $C(\mathbb{T}^d)$, recovering the universal approximation property.

% In particular, the requirement that $L\to\infty$ for universal approximation holds even for datasets that have a finite degree of precision. This will be important in the next sub-section where we analyze the approximation power of bit-bit-encoding.

\section{Techniques \& Results: Bit-bit Encoding Schemes}
Bit-bit encoding maps numeric features and class labels into bitstrings for quantum classification. A necessary aspect of this process is compressing the original dataset. In this section, we empirically investigate how many qubits are required to represent bit-bit encoded datasets for the purpose of the learning task. We consider training and testing sets separately when calculating coverage. We calculate the training set coverage as $1 - \text{training collision incidence}$, where the training collision incidence is the proportion of training samples with different class labels which have the same encoding. A training set is fully covered when there are no collisions between samples with different class labels, that is, all samples with the same encoding have the same class label. In this case, since there are no colliding samples with different classes, a perfect model would theoretically be able to achieve $100\%$ training accuracy. Therefore, we refer to $1 - \text{training collision incidence}$ as the theoretical training accuracy. We define the testing set coverage as $1 - \text{testing overlap incidence}$, where the testing overlap incidence is the proportion of testing samples which overlap with the training data but do not have the same class label as the majority class label of the overlapping training data. A testing set is fully covered when all of the samples either 1.) are not overlapping with the training data or 2.) have the same class label as the majority class label of the overlapping training data. We refer to $1 - \text{testing overlap incidence}$ as the theoretical testing accuracy. We define the number of qubits at which the training and testing sets are covered (i.e. $100\%$ theoretical accuracy) as $Q_{\text{train}}(1.0)$ and $Q_{\text{test}}(1.0)$, respectively. The number of qubits required to cover the class labels is $Q_y = \lceil \log_2{(\text{number of classes})} \rceil$. Then, the number of qubits to cover an entire dataset is $Q_{\text{dataset}}(1.0) = \max{(Q_{\text{train}}(1.0), Q_{\text{test}}(1.0))} + Q_y$. The number of qubits required to reach an arbitrary accuracy $x$ is defined as $Q_\text{dataset}(x) =\max{(Q_{\text{train}}(x), Q_{\text{test}}(x))} + Q_y$. 

While in the bit-bit encoding algorithm, the number of components after dimensionality reduction, $D$, is a modifiable parameter, unless noted otherwise, we set $D$ equal to the original number of features.

\subsection{Benchmark Datasets with Principal Component Analysis} \label{sec:pca}
% MNIST example
\begin{figure}
    \centering
    \includegraphics[width=\linewidth]{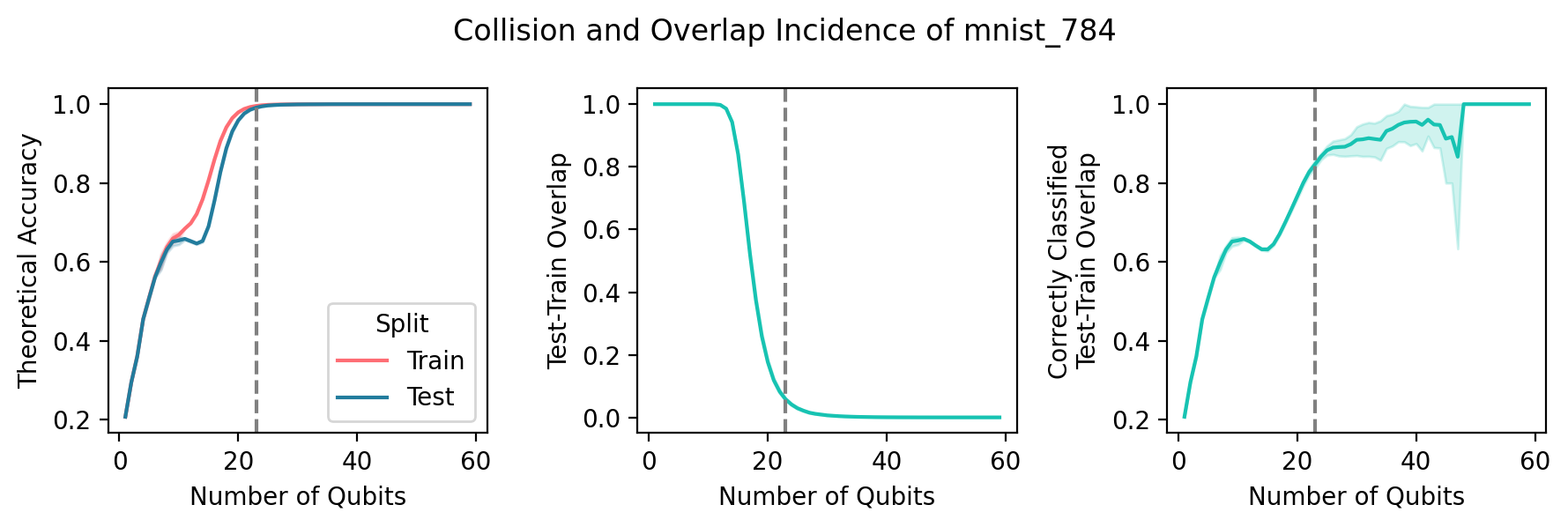}
    \caption{The train and test theoretical accuracy (left), test-train overlap (middle), and correctly classified test-train overlap (right) over the number of allocated qubits for the MNIST$\_784$ OpenML dataset with PCA dimensionality reduction. Each line shows the mean of the corresponding metric with $95\%$ confidence interval error bands. The gray dotted line is the mean $Q_\text{dataset}(0.99)$ (minus the class qubits). In the rightmost plot, the dip in the testing accuracy between $10$ and $20$ qubits is caused by shifts in the distribution of bits allocated to the individual features. Between $20$ and $50$ qubits, test-train overlap is near but not quite $0$, so the small number of overlapping samples lead to greater shifts in the overall proportion of correctly classified samples.}
    \label{fig:mnist}
\end{figure}
We start with a simple example of calculating $Q_\text{dataset}(x)$ for the MNIST digit-recognition dataset. We use the MNIST$\_784$ dataset from OpenML \cite{openml}, which flattens the $28$x$28$ images into $784$ features, where the features are the grayscale intensity of each pixel. The MNIST dataset contains $70,000$ samples and $10$ classes, one class for each single-digit number. Analyzing the behavior of bit-bit encoding on this dataset, as we do below, illustrates how the encoding can already indicate the achievable accuracy and generalization performance of a quantum model before it is trained.

Figure~\ref{fig:mnist}, left, shows the mean theoretical accuracy across $10$ replicates of train/test splits. The maximum $Q_\text{dataset}(1.0)$ for MNIST $\_784$ is $63$, where the x-axis of the figure is cut-off, while the minimum $Q_\text{dataset}(1.0)$ is $42$ and the mean is $50$. The gray dotted line shows the mean $Q_\text{dataset}(0.99)$, $27$, minus the number of class qubits, $4$. By subtracting the class qubits, the gray line marks the average $\max{(Q_{\text{train}}(0.99), Q_{\text{test}}(0.99))}$ (recall that the theoretical train accuracy is the same as $Q_{\text{train}}(x)$, similarly for test). We postulate that large difference between $Q_\text{dataset}(1.0) = 63$ and $Q_\text{dataset}(0.99) = 27$ reveals a trade-off between achieving the maximum theoretical training accuracy and obtaining a generalized model. We see in the middle plot of Figure~\ref{fig:mnist} that as the number of qubits increases, the overlap between the training and testing sets decreases. Between $Q_\text{dataset}(0.99)$ and $Q_\text{dataset}(1.0)$, specifically, the test-train overlap is near but not quite zero. Thus, between $Q_\text{dataset}(0.99)$ and $Q_\text{dataset}(1.0)$, the discretization of the MNIST components are becoming more and more refined for only a few samples, which reduces the overall generalizability of the encoding to new samples, or in other words, is overfitting. Furthermore, as seen in Figure~\ref{fig:mnist}, right, as the number of qubits increase, the test samples that do overlap with train increasingly have the same class label as the majority label of the train samples. Between $Q_\text{dataset}(0.99)$ and $Q_\text{dataset}(1.0)$, there are so few samples overlapping between test and train that slight shifts in the encoding causes the region of stochasticity in the plot.

% Methods + Results: OpenML datasets + Q_dataset calculation
Next, we calculate $Q_\text{dataset}$ for a suite of benchmark machine learning datasets \cite{feurerauto2021}. We filter the benchmark datasets to those that fulfill the following criteria: less than $1,000$ features which are all numeric, less than $10,000$ samples, more samples than features, no missing values, single-label classification with less than $10$ classes, and no collisions when not encoded. After filtering, we have $28$ datasets. Supplementary Figure S1 visualizes the number of samples, number of features, and number of class labels of each dataset. We randomly split each dataset into $80\%$ training and $20\%$ testing, repeated across $10$ replicates. We find the mean $Q_\text{dataset}$ for the benchmark datasets to be $27.02$ qubits, $95\%$ CI $[26.11, 27.94]$. Supplementary Figure S2 shows the theoretical accuracy as the allocated bits increase for the training and testing sets. We see each dataset converges to its $Q_\text{dataset}(1.0)$ at a different rate and that there is minimal stochasticity in the theoretical accuracy across replicates.

% Ablation study
We also investigate the effect of different bit-bit encoding parameters under PCA on $Q_\text{dataset}(1.0)$. In Supplementary Figure S3, we see that without the copula transform step, the mean $Q_\text{dataset}(1.0)$ of the benchmark datasets is $40$, while with the copula transform, it is $27$. The copula transform spreads samples more evenly across the discrete bins, so fewer cross-class samples fall into the same encoded bitstring and fewer test samples fall into training encodings with a conflicting majority label. In Supplementary Figure S4, we see that the mean $Q_\text{dataset}(1.0)$ of the benchmark datasets is largely unaffected by setting the number of components, $D$, to a percentage of the true number of features. Yet, in Supplementary Figure S5, we see whether the mean $Q_\text{dataset}(1.0)$ of individual datasets is higher or lower with fewer PCA components varies. On one hand, OpenML dataset ``wall-robot-navigation'''s mean $Q_\text{dataset}(1.0)$ is $42$ bits higher with $24$ components (the number of features) than with $4$ components ($80\%$ reduction). On the other hand, OpenML dataset ``steel-plates-fault'''s mean $Q_\text{dataset}(1.0)$ is $20$ bits lower with $33$ components (the number of features) than with $26$ component ($80\%$ reduction). We hypothesize that datasets with redundant or low-quality features exhibit lower $Q_\text{dataset}(1.0)$ with a smaller dimensionality, while datasets where class information is distributed across many low-variance directions will exhibit higher  $Q_\text{dataset}(1.0)$ with a smaller PCA dimensionality.

\subsection{Benchmark Datasets Across Dimensionality Reduction Schemes} \label{sec:alternatives}
% Methods: dimensionality reduction schemes
We now compare $Q_\text{dataset}(1.0)$ for the benchmark datasets \cite{feurerauto2021} across different dimensionality reduction schemes. We include a null comparison where we don't transform the data with a dimensionality reduction scheme (NONE). We compare three linear schemes: PCA, Independent Component Analysis (ICA), and truncated singular vector decomposition / latent semantic analysis (LSA). We also compare two non-linear schemes: radial basis function kernel PCA (KPCA) and uniform manifold approximation and projection (UMAP). All schemes, except UMAP \cite{umap}, are implemented with Scikit-Learn \cite{scikit-learn}. Some dimensionality reduction schemes require special processing of the data --- for KPCA, the training and testing sets are min-max normalized before reduction, and for ICA, the number of components is set to the minimum between the number of features and the rank of the training data.

% Results: Q_dataset comparison across schemes
\begin{figure}
    \centering
    \includegraphics[width=0.6\linewidth]{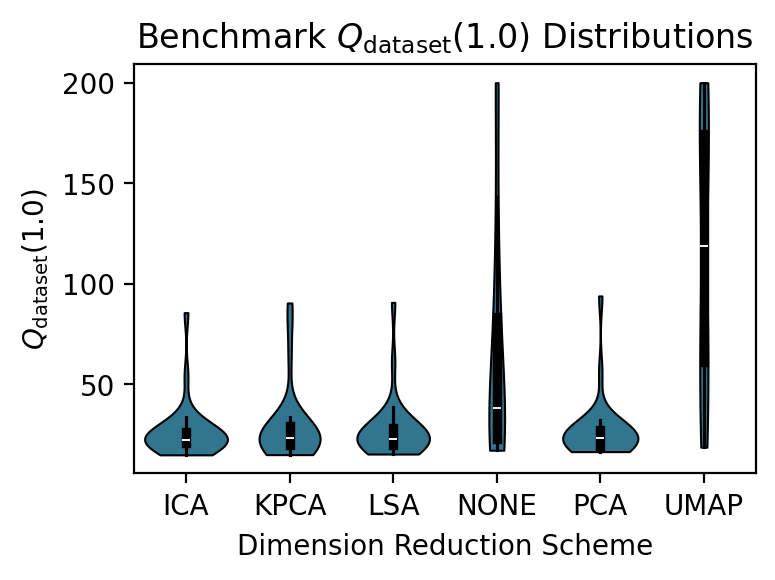}
    \caption{Violin plots of $Q_\text{dataset}(1.0)$ for OpenML benchmark datasets across dimensionality reduction schemes. Each violin depicts the probability density of $Q_\text{dataset}(1.0)$ from the mean $Q_\text{dataset}(1.0)$ of each benchmark dataset.}
    \label{fig:alternatives}
\end{figure}
Figure~\ref{fig:alternatives} shows the distributions of $Q_\text{dataset}$ across dimensionality reduction schemes. Overall, ICA has the lowest rounded mean $Q_\text{dataset}$ at $26$, followed by PCA ($27$), LSA ($28$), KPCA ($29$), NONE ($51$), and UMAP ($90$). The long tails are due to an outlier dataset in the suite of benchmark datasets --- the ``wall-robot-navigation'' OpenML dataset \cite{openml}. While each other dataset in the suite has a $Q_\text{dataset}(1.0) < 50$ under PCA dimensionality reduction, ``wall-robot-navigation'' has a mean $Q_\text{dataset} = 94$. This indicates significant overfitting is required to obtain full accuracy on this dataset, likely due to a few difficult-to-classify samples; for further discussion, see Section~\ref{sec:rore}.

PCA, LSA, and KPCA transform the original feature space into uncorrelated orthogonal components, while ICA produces statistically independent components. Independent features are vital for the bit allocation step of bit-bit encoding. To see this, let $X = [x_1, \dots, x_D]$ denote the $D$ components after dimensionality reduction and $y$ the class labels. Bit-bit encoding allocates bits to a component $i$ proportional to $I(X_i;y)$, the mutual information between component $i$ and class labels $y$. If the components are not independent, they will have overlapping information, i.e. $I(X_i;X_j) > 0$, which is over-counted in the bit allocation. The over-count can be written as $\sum_{i=1}^D{(I(X_i;y))}-I(X;y)$, where $I(X;y)$ is the joint mutual information between the set of components and the class. Note that even with independent components from dimensionality reduction schemes such as PCA, there still may be unoptimal distribution of bits to components, meaning that $Q_\text{dataset}(x)$ is not a theoretical lower bound. The unoptimal bit distribution comes from dimensionality reduction schemes not making the components independent with regards to the class. The components may appear to be independent when not considering the class ($I(X_i;X_j) = 0$), but in reality have synergist information with the class ($I(X_i;X_j;y) > 0$) and redundant information with each other ($I(X_i;X_j|y) < 0$) or vice versa \cite{timme2014synergy}.

\subsection{Results on Resource Estimation} \label{sec:rore}
\begin{figure}
    \centering
    \includegraphics[width=0.6\linewidth]{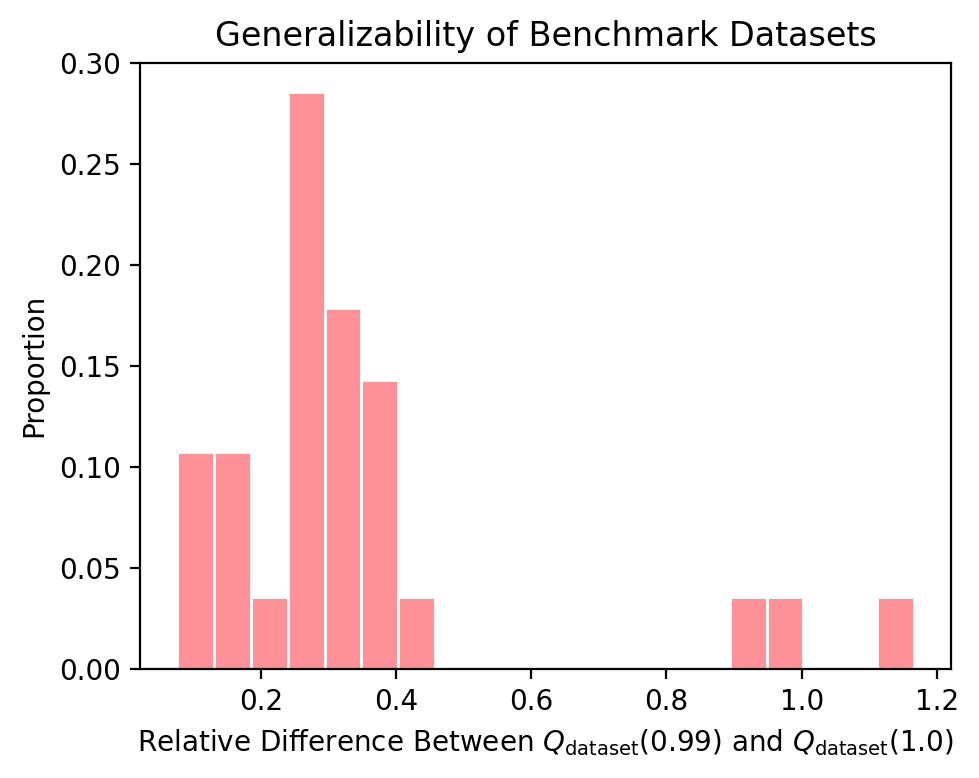}
    \caption{The pairwise relative differences between $Q_\text{dataset}(0.99)$ and $Q_\text{dataset}(1.0)$, $\frac{Q_\text{dataset}(1.0)-Q_\text{dataset}(0.99)}{Q_\text{dataset}(0.99)}$, for each benchmark dataset.}
    \label{fig:generalization}
\end{figure}

% Generalizability
In Section~\ref{sec:pca} we saw a dramatic difference in $Q_\text{dataset}(0.99)$ and $Q_\text{dataset}(1.0)$ for the MNIST dataset. Figure~\ref{fig:generalization} shows the relative difference between $Q_\text{dataset}(0.99)$ and $Q_\text{dataset}(1.0)$ for each benchmark dataset \cite{feurerauto2021}. The three datasets with the highest difference are ``wall-robot-navigation'', ``Satellite'', and ``GesturePhaseSegmentationProcessed'' \cite{openml}. In these cases, ambiguous, noisy, or contradictory samples require substantially more qubits to distinguish. A higher difference suggests that significant over-fitting would be required to train a model to $100\%$ accuracy on those datasets, highlighting a trade-off between model complexity and resource requirements, which should be considered when applying quantum machine learning as additional qubits may yield a decreased ability for a model to generalize.

% Quantum advantage
% My understanding to the justification of the 50 number:
% A classical simulation of a quantum circuit with N qubits requires storing 2^N floats
% In 2019, Google showed a limit in classical simulation at ~50 qubits (“Quantum supremacy using a programmable superconducting processor”, Frank Arute et al., 2019)
% Earlier this year, Quantinuum showed that the difficulty of classically simulating their H2 quantum computer is limited only by the number of qubits (“Computational Power of Random Quantum Circuits in Arbitrary Geometries”, Matthew DeCross et al., 2025)
% Thus, we say that requiring >50 qubits is an indicator of potential quantum advantage
The universal approximation property of bit-bit encoding allows us to state that a dataset can be completely modeled with $Q_\text{dataset}(1.0)$ qubits. When entanglement is not bounded, classical simulation capabilities, including those that use tensor networks, plateau at around 50 qubits \cite{quantinuum_rcs_2025}. Hence, we can claim that if $Q_\text{dataset}$ is $\lessapprox 50$ qubits, the corresponding dataset is unlikely to benefit from quantum advantage. Datasets that require $\gtrapprox 50$ qubits are promising targets for finding quantum advantage in machine learning within this framework. In Section~\ref{sec:pca}, we saw that classical machine learning datasets encoded to minimize statistical dependency between features had a mean $Q_\text{dataset}(1.0) = 27$. These classical machine learning datasets had less than $1,000$ features, less than $10,000$ samples, and less than $10$ classes (Supplementary Figure 1). This result suggests that classic, medium-sized machine learning datasets are unlikely to benefit from quantum advantage.

\subsection{Large Transcriptomic Dataset} \label{sec:tahoe}
\begin{figure}
    \centering
    \includegraphics[width=0.35\linewidth]{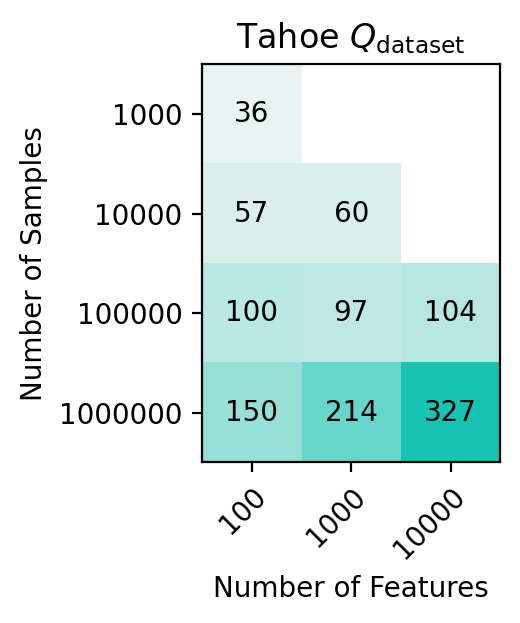}
    \caption{A heatmap visualizing how the rounded mean $Q_\text{dataset}(1.0)$ of subsamples of the Tahoe dataset scale with the number of features (x-axis) and number of samples (y-axis).}
    \label{fig:tahoe_scale}
\end{figure}

% Motivation
The default bit-bit encoding algorithm requires the entire dataset to be read into memory, which becomes impractical for large datasets. Furthermore, dimensionality reduction algorithms may not scale well as the dataset size increases, leading to intractable time and memory overhead. For example, PCA dimensionality reduction scales polynomially with dataset size. Learning problems on large datasets are more likely to benefit from quantum advantage, since large datasets should have more  information that can only be encoded without collision with a significant number of qubits. Therefore, we extend the bit-bit encoding algorithm to operate in a batched streaming mode, enabling the processing of large-scale datasets. 

% Methods: batching
Batched bit-bit encoding begins by streaming the training data in batches of a specified size and incrementally fitting the dimensionality reduction algorithm. After all training batches have been processed and the dimensionality reduction model has been fit, the training data are streamed a second time. In this pass, each batch is encoded using the fitted dimensionality reduction scheme, after which batch-level mutual information importance scores are computed and the minimum and maximum values for each feature are recorded. The overall feature importance scores are obtained by averaging the importance scores across all batches. In a third streaming pass, a bin-based copula transform model is fit on the data. In a final streaming pass, the training data are re-encoded, min–max scaled, copula transformed, and written to disk along with the fitted dimensionality reduction model, feature importance scores, copula model, and per-feature minimum and maximum values. The testing data are then streamed and transformed using the saved models and statistics. To compute $Q_\text{dataset}(x)$, both training and testing batches are discretized according to the specified number of qubits and the saved feature importance scores. As the training and testing batches are streamed, unique bit strings, the counts of class labels associate with each bit string, and a count of the total number of samples are stored in memory, allowing collision incidences to be computed in the same manner as in the non-batched setting.

% Methods: Tahoe dataset
We study how $Q_\text{dataset}(1.0)$ scales with both the number of features and the number of samples in a single large dataset. To this end, we subsample from the Tahoe-100M dataset \cite{tahoe}, which comprises approximately $100$ million transcriptomic profiles collected across $50$ cancer cell lines and $379$ drugs at varying dosages (corresponding to $1,100$ small-molecule perturbations). Each of the $100$ million samples is represented by $62,710$ features (gene expression measurements). For our analysis, we restrict the dataset to a single cell line, NCI-H295 (RRID:CVCL\_0456), which is the most frequently seen in the Tahoe dataset at $6\%$ of the samples. We formulate the corresponding prediction task as a multiclass classification problem: given the gene expression profile of a sample, identify which drug was applied to the NCI-H295 cell line. After filtering, the resulting Tahoe subset contains $6,040,372$ samples, $52,783$ features, and $65$ classes (the drugs applied to the NCI-H295 cell line, specifically). We then further subsample from the established Tahoe subset so we can investigate how $Q_\text{dataset}(1.0)$ scales with the number of features and number of samples. We subsample both the number of samples and the number of features logarithmically, keeping the number of features smaller than the number of samples. We replicate each subsample $3$ times with a random train-test split and a biased feature selection. The feature selection is biased such that genes are chosen as features with a probability proportional to their frequency in the Tahoe subset. This is because gene expressions have been found to exhibit a power-law distribution \cite{gene_powerlaw}, so the biased feature sampling prevent excessive sparsity in the subsample. Samples are subsampled progressively --- for example, the subsamples with $1000$ samples will contain all the samples from the $100$ sample subsample. We obtain $Q_\text{dataset}(1.0)$ by calculating the training and testing collision incidence when continuously increasing the number of qubits allocated to the subsample in increments of $10$ until either the training and testing collision are both $0$ or we reach $400$ qubits. We use batched bit-bit encoding to calculate $Q_\text{dataset}(1.0)$, where the batch size is set to the number of samples divided by $10$.

% Results: scale of Q_dataset across dataset size
Figure~\ref{fig:tahoe_scale} shows the mean $Q_\text{dataset}(1.0)$ over $3$ replicates as the number of features and number of samples increase. We see that $Q_\text{dataset}(1.0)$ scales with the number of samples, and below $1,000,000$ samples, $Q_\text{dataset}(1.0)$ stays constant with the number of features. At less than $1,000,000$ samples, the number of samples is smaller than the effective number of distinct bit strings determined by the allocated bits, so collisions are rare and $Q_\text{dataset}(1.0)$ is largely insensitive to the number of features. At $1,000,000$ samples, $Q_\text{dataset}(1.0)$ scales with the number of features; collisions begin to appear because it is more likely for multiple samples to map to the same bit strings. In this case, adding more low-information features reduces the bits allocated to informative features, increasing collisions, whereas a smaller set of highly informative features receives more bits per feature, allowing samples to be encoded more uniquely and reducing collisions. This result emphasizes the importance of feature selection to minimize the influence of low-information features in the encoding.

 % Results: compare to benchmark datasets
In Section~\ref{sec:pca}, we saw that benchmark machine learning datasets, which had $<10,000$ samples and $<1000$ features, had a mean $Q_\text{dataset}(1.0) = 27$. We see the Tahoe subsamples at the same dimensionality have a higher mean $Q_\text{dataset}(1.0) = 51$. The transcriptomic subsets requiring more qubits than similarly-sized benchmark datasets reflects a growing pattern in which quantum machine learning is emerging as a potentially powerful tool for addressing the unique complexity present in biological learning problems \cite{emani2021quantum}, \cite{ kubal2025quantumplatformmultiomicsdata}.

\section{Training}
\begin{figure}
    \centering
    % First plot
    \begin{subfigure}{0.45\textwidth}
        \centering
        \includegraphics[width=1.1\linewidth]{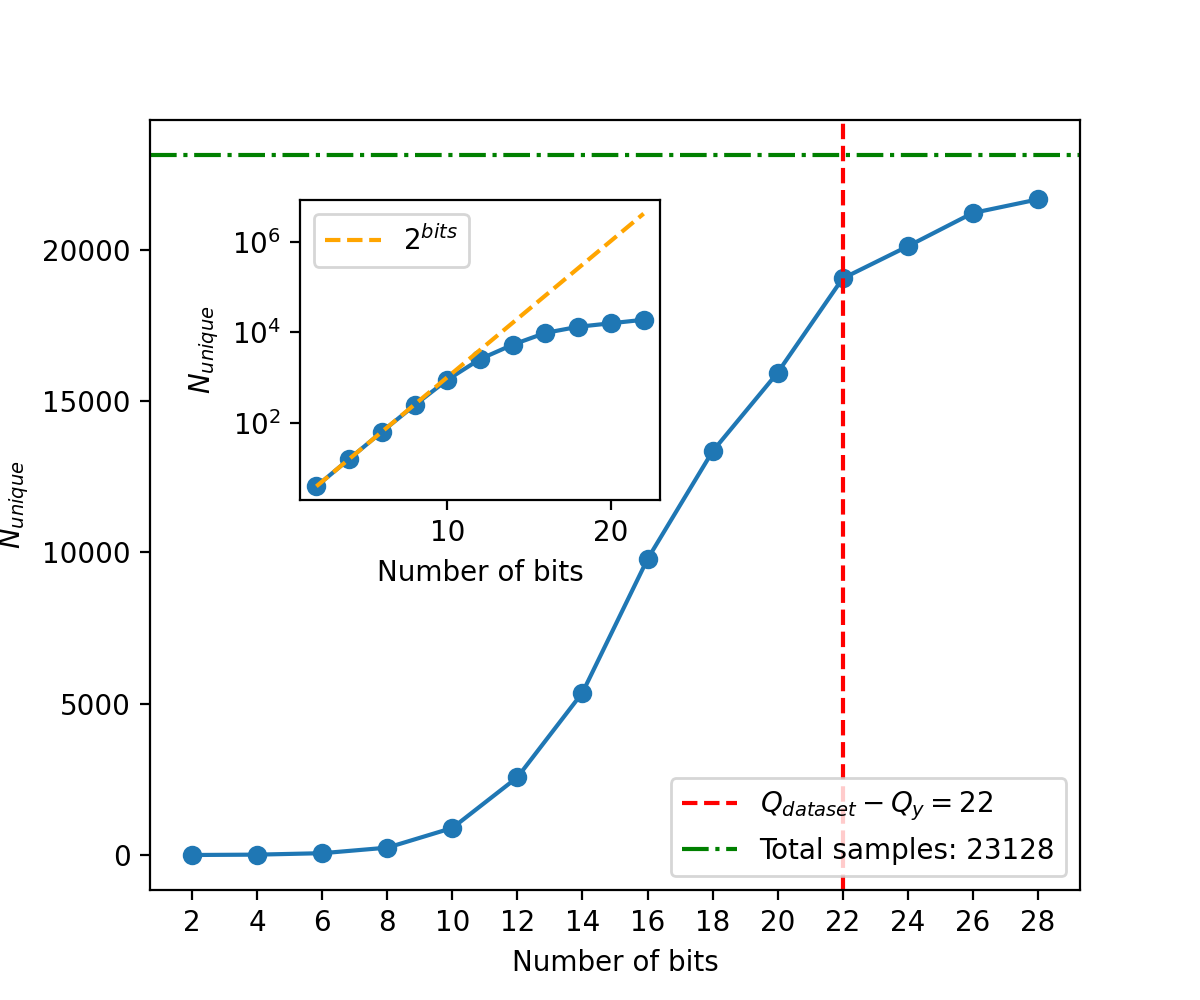}
    \end{subfigure}
    \hfill
    % Second plot
    \begin{subfigure}{0.45\textwidth}
        \centering
        \includegraphics[width=1.1\linewidth]{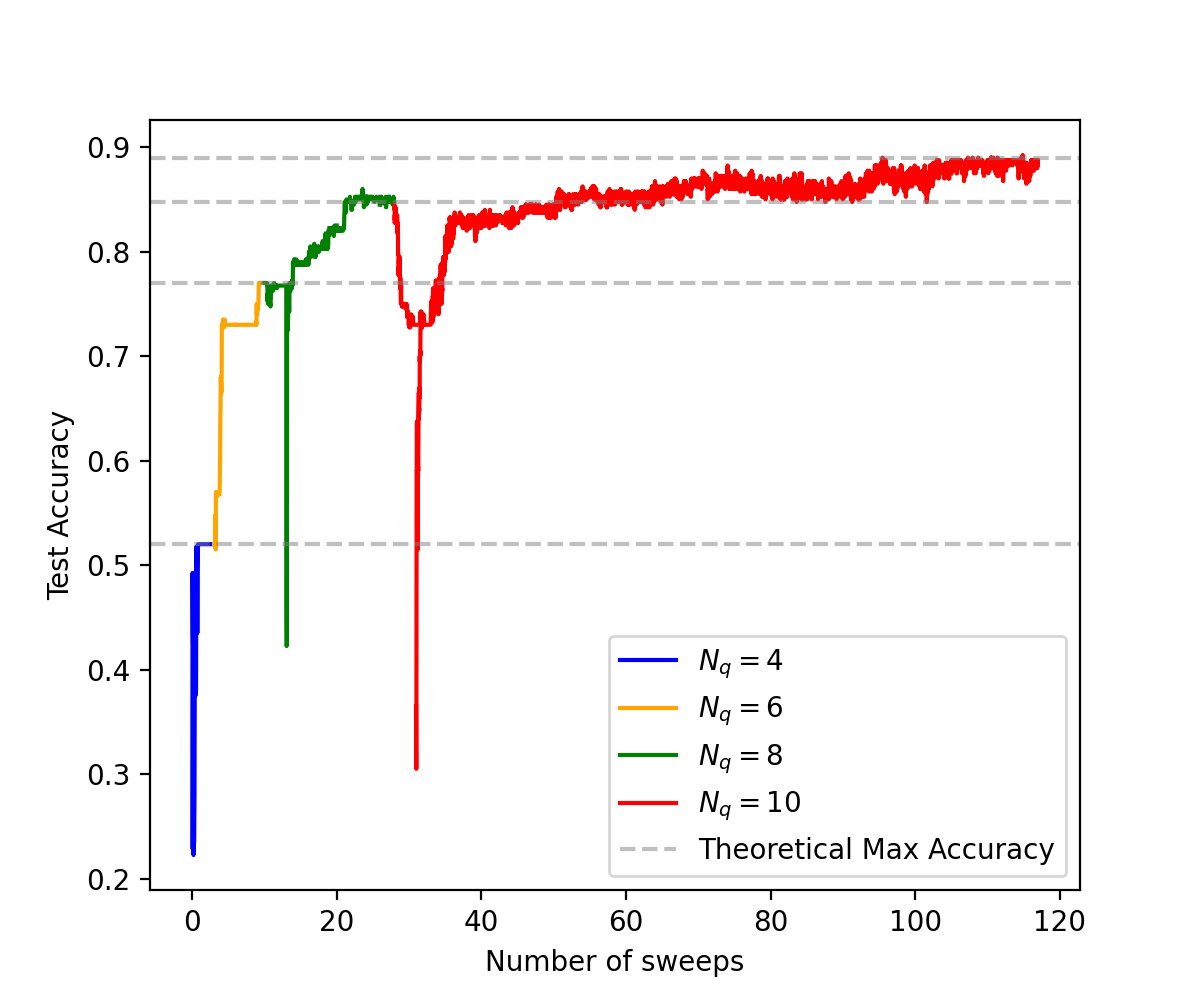}
    \end{subfigure}
    \caption{(left) The number of unique samples as a function of the number of bits used to encode the input. The inset shows the same curve with a log scale on the y-axis and compares it with an exponential. (right) Evolution of test accuracy for the first 4 digits of MNIST. $N_q$ is the number of qubits. 2 qubits are reserved for reading out the class label, and $N_q-2$ qubits are available to load the data sample at the input.}
    \label{fig:training}
\end{figure}

In quantum machine learning, models that are expressive are often associated with barren plateaus and thus thought to be difficult to train \cite{mcclean2018barren}. While this paper focuses on creating a framework for calculating the number of qubits required for a dataset, we also show a brief demonstration of training of a quantum model with bit-bit encoding. We use a dataset consisting of first 4 digits of MNIST. We use the model architecture, sub-net initialization, and exact coordinate update optimization techniques from \cite{johri2025bitbitencoding}. We create a restricted training set by removing the collisions at each model size. The training batches consist of only the unique samples that occur after collision removal. This is possible as the number of samples grows slower than exponentially with the number of bits as shown in Figure~\ref{fig:training} (left). Interestingly, even when the dataset is fully encoded, the number of unique samples in the encoded dataset is less than the total number of samples in the dataset.

% \begin{figure}
%     \centering
%     \includegraphics[width=0.6\linewidth]{figures/sub_net_chain_performance-4-10q.png}
%     \caption{Evolution of test accuracy for the first 4 digits of MNIST. $N_q$ is the number of qubits. 2 qubits are reserved for reading out the class label, and $N_q-2$ qubits are available to load the data sample at the input.}
%     \label{fig:training}
% \end{figure}

Figure~\ref{fig:training} (right) shows the evolution of test accuracy as a function of the number of training sweeps for models with different numbers of qubits $N_q$. In the exact coordinate update optimizer, one parameter is updated at a time, and a sweep denotes one full pass through all parameters. The dashed horizontal lines indicate the theoretical maximum accuracy achievable at that number of qubits. As can be seen, all the models converge to this maximum accuracy predicted by bit-bit encoding. Larger models require more sweeps to converge. They also exhibit transient drops in accuracy due to the use of higher-order loss functions which are more unstable during training.

\section{Discussion and Future Directions}
In this paper, we examined the universal approximation and resource estimation characteristics of bit-bit encoding for loading classical data into quantum computers. Naively, one may assume that encoding in the basis is useful only when the input data is naturally in the form of binary strings, for example, binary images, combinatorial optimization problems, or molecular orbital occupation vectors. Contrary to this intuition, we have demonstrated that datasets with real-valued entries can also be efficiently represented as bit strings.

We proved that bit-bit encoding inherently ensures universal approximation. Moreover, for a fixed number of qubits, the achievable accuracy can be efficiently determined from the dataset itself, without requiring model training. Taken together, these results establish the first systematic framework for resource estimation in quantum machine learning.

Other encoding schemes have been previously explored with varying degrees of success, but their expressive power is far less predictable, particularly when classical preprocessing and compression are involved. As a result, they lack a systematic framework for resource estimation and provide no reliable way to judge whether the same outcomes could be achieved with fewer qubits. This makes it nearly impossible to meaningfully assess quantum advantage. While such methods may appear to work on small datasets, without a principled understanding of their expressivity, they cannot be expected to scale. Many previous quantum machine learning studies have also leaned on hybrid models, with part of the model parameterized classically and part quantum, which further blurs the line of advantage. In sharp contrast, our model relies exclusively on quantum parameters, making the presence and scope of quantum advantage in learning far clearer.

An additional advantage of bit-bit encoding is that it enables an intrinsic evaluation of quantum model performance, independent of comparisons to classical baselines, since the maximum achievable accuracy is known a priori. Conventionally, the performance of quantum models is typically judged relative to classical models; however, classical accuracy can vary widely and may even depend on the expertise of the practitioner training the model. By contrast, bit-bit encoding offers an objective criterion for assessing how well a quantum model has learned.

In the numerical sections of the paper, we applied bit-bit encoding to various real datasets and investigated its performance and dynamics. While we cannot theoretically prove that we have an optimal encoding, if the dimensionality reduction scheme produces independent features, it is close to optimal in an information theoretical sense. As classical simulation hits a wall around $50$ qubits, we can thus state with moderate confidence that if a dataset can requires more than about $50$ qubits to encode, it is a candidate for quantum advantage within this framework. In particular, it is notable that medium-sized benchmark datasets can be fit within $27$ qubits using our method, which implies they will not go beyond classical simulation capabilities. This, in turn, indicates that several datasets that have been studied in the quantum literature so far may not be candidates for quantum advantage. When applying a batched version of bit-bit encoding to a large-scale transcriptomic dataset, we saw the importance of using a dataset with a comprehensive set of samples and informative features. Too many uninformative features will cause bit-bit encoding to estimate a larger $Q_\text{dataset}(x)$ than necessary as the PCA and mutual information steps of bit-bit encoding struggle to deal with the noise.

For future work, we plan to extend bit-bit encoding to convert more data types into quantum basis states, including sequential, multi-label, regression, and spatial. We also plan to focus on investigating the potential for quantum advantage for datasets from specific fields, such as biology and chemistry. We will aid this investigation by validating $Q_\text{dataset}(x)$ with training experiments that compare the scaling of analogous classical and quantum models.

While this work has primarily focused on determining the number of qubits required for encoding, many open directions remain in developing a deeper understanding of bit-bit encoding. One important question concerns the optimal number of qubits for generalization: although increasing qubits improves precision, it may also introduce a trade-off between perfectly fitting the training data and effectively generalizing to unseen examples. Another open challenge is to move beyond logical qubits and identify what ansatz structures yield efficient models when implemented on physical qubits. Finally, questions remain about the gate complexity and learning dynamics of models trained on bit-bit encoded data. These issues will be the focus of future investigations.

Finally, in addition to the qubits required for encoding, a full assessment of quantum advantage for a dataset requires analyzing quantities that measure the entangling power, such as operator Schmidt rank, induced multipartite correlation functions, and metrics of non-classicality such as negativity or contextuality, of the classification unitary operator. Upcoming work will compute these quantities to establish a more complete framework for understanding quantum advantage in machine learning.

\section{Software Framework}
The techniques in the paper are implemented in Red Cedar, a software framework being developed at Coherent Computing Inc. It can be made available upon request.

\bibliography{references}

% \documentclass{article}
% \usepackage{graphicx} % Required for inserting images
% \usepackage[square,numbers]{natbib}
% \usepackage{amsmath}
% \usepackage{xcolor}
% \usepackage{braket}

% \begin{document}

\section*{Supplemental Figures}

\begin{figure}[h!]
    \centering
    \includegraphics[width=0.75\textwidth]{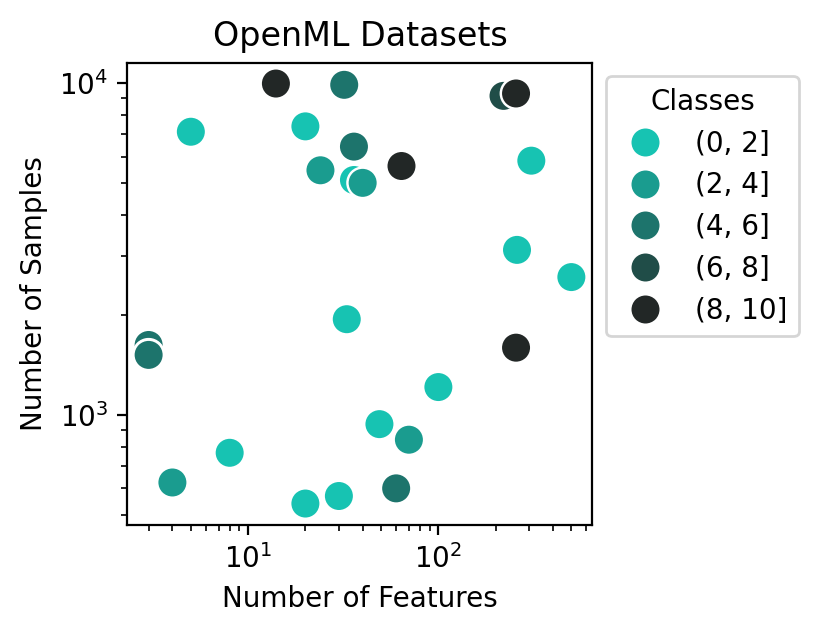}
    \caption{The number of features (x, log-scale), number of samples (y, log-scale), and number of classes (hue) of each of the $28$ benchmark datasets used in experiments for Section 4.1 and Section 4.2.}
    \label{fig:datasets}
\end{figure}

\begin{figure}
    \centering
    \includegraphics[width=0.49\linewidth]{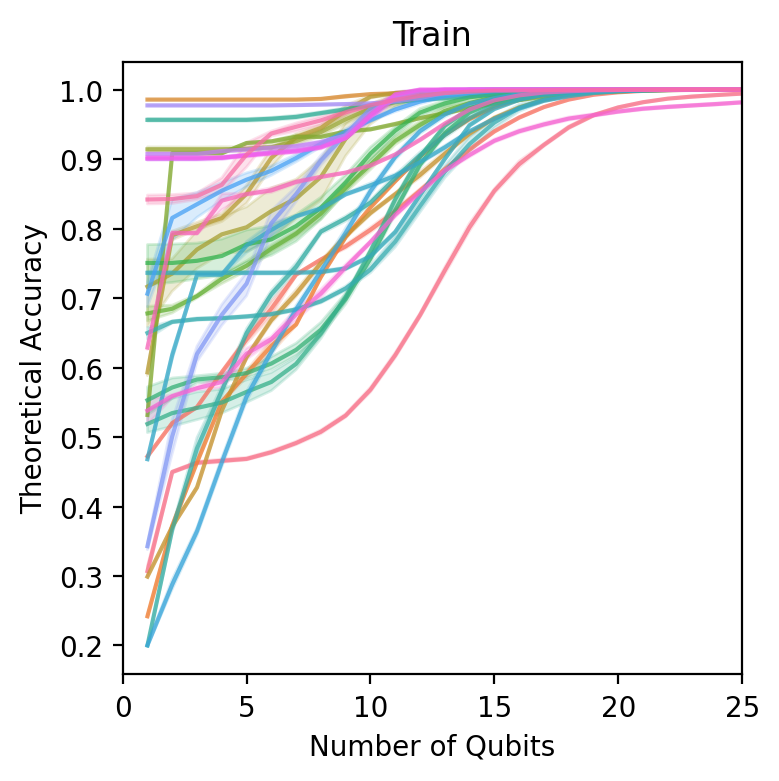}
    \includegraphics[width=0.49\linewidth]{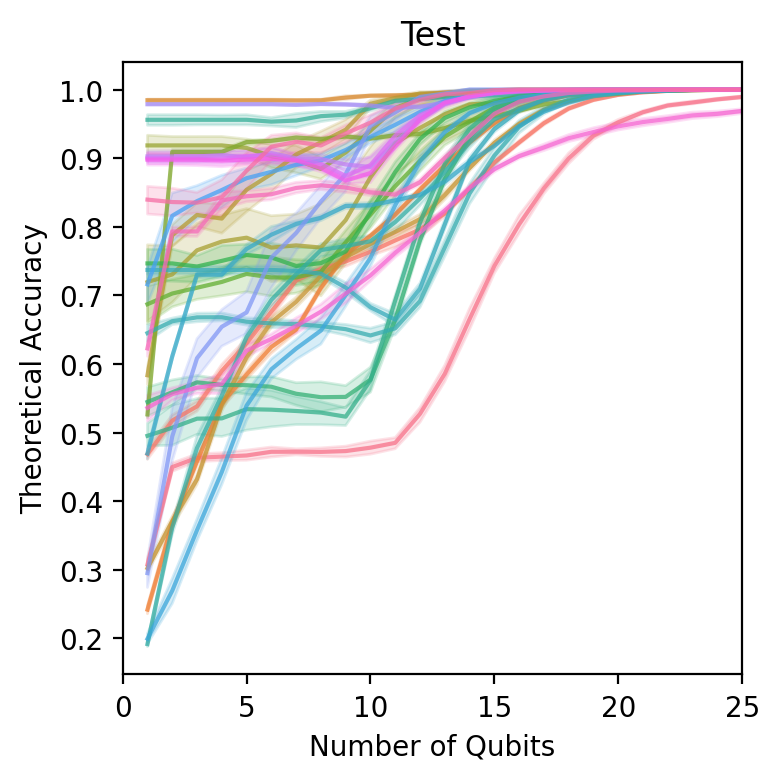}
    \caption{The theoretical training and testing accuracy across allocated qubits for the experiment with PCA dimensionality reduction. Each line is the mean for each dataset, with $95\%$ confidence interval error bands. The x-axis is cut off at $25$ qubits for increased visibility of individual dataset trends.}
    \label{fig:pca_collision_incidence}
\end{figure}

\begin{figure}
    \centering
    \includegraphics[width=0.75\textwidth]{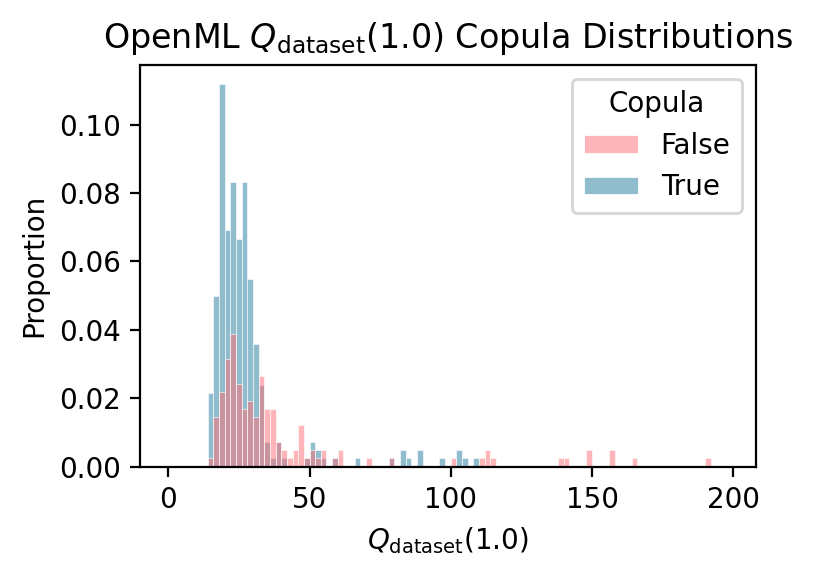}
    \caption{The difference in mean $Q_\text{dataset}(1.0)$ distributions of the benchmark datasets when running bit-bit encoding with and without the copula transform.}
    \label{fig:copula}
\end{figure}

\begin{figure}
    \centering
    \includegraphics[width=0.75\textwidth]{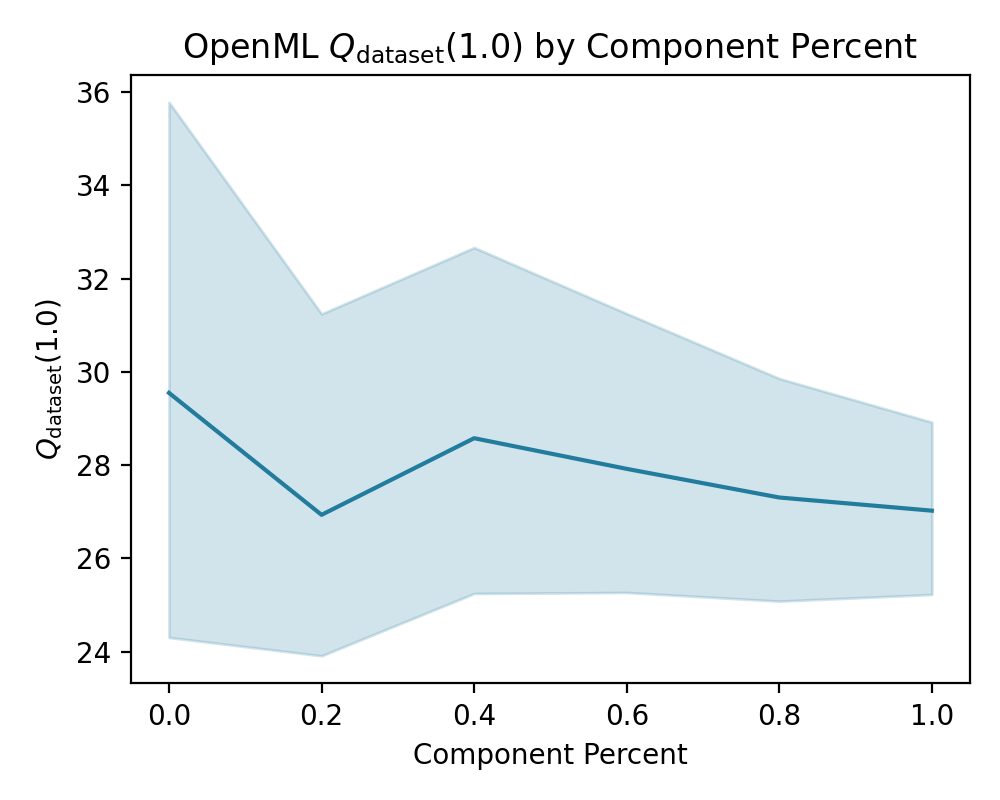}
    \caption{The mean $Q_\text{dataset}(1.0)$ of the benchmark datasets when running bit-bit encoding with different number of PCA components. The number of components is given by the floor of the number of features multiplied by the component percent.}
    \label{fig:qdataset_by_dim}
\end{figure}

\begin{figure}
    \centering
    \includegraphics[width=0.75\textwidth]{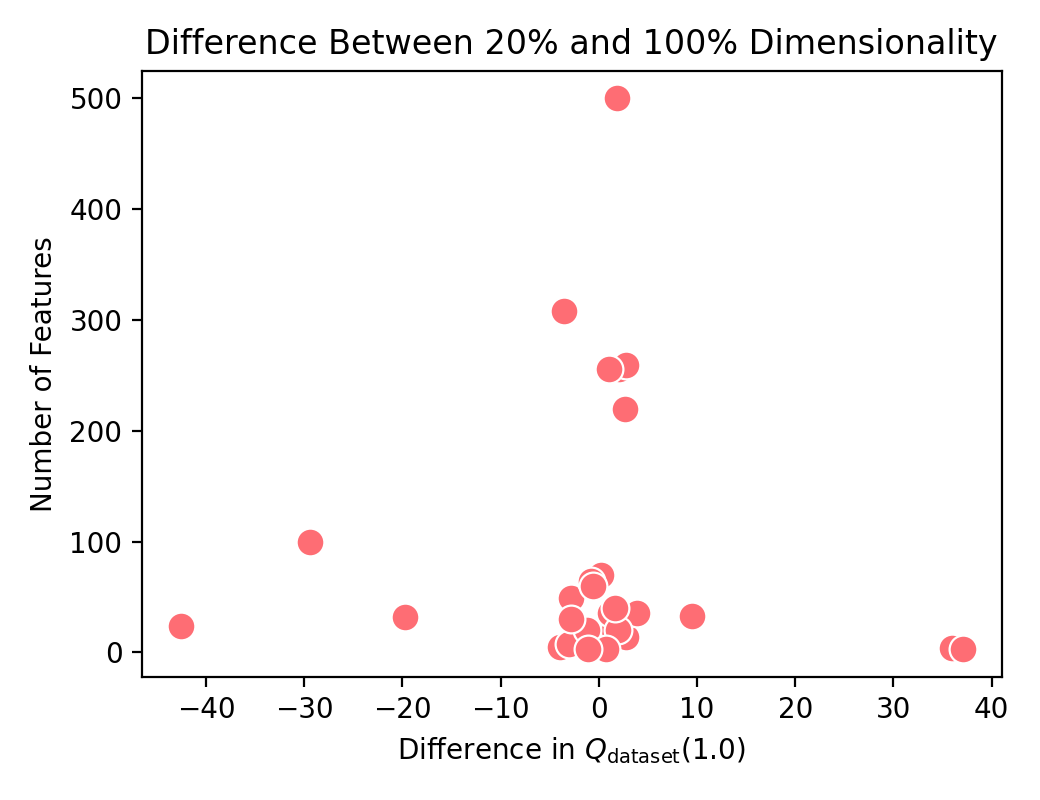}
    \caption{The difference between mean $Q_\text{dataset}(1.0)$ for each of the benchmark datasets when running bit-bit encoding at $20\%$ component percent and $100\%$ component percent.}
    \label{fig:dim_by_feature}
\end{figure}

% \footnotesize
% \bibliographystyle{naturemag}
% \bibliography{references}

\end{document}